\documentclass[12pt]{iopart}
\usepackage{tikz}
\usepackage{makeidx,amssymb,amsthm,graphicx}
\usepackage[toc,page]{appendix}
\usepackage{iopams}  
\begin{document}

\def\neel{{N\'eel} }


\title{Magnetic anisotropy in HoIr$_2$Si$_2$ ($I4/mmm$)}

\author{K.Kliemt}
\ead{kliemt@physik.uni-frankfurt.de}
\address{Physikalisches Institut, Goethe-Universit\"at Frankfurt/M, 60438 Frankfurt/M, Germany}
\author{M.Bolte}
\address{Institut f\"ur Anorganische und Analytische Chemie, Goethe-Universit\"at Frankfurt/M, 60438 Frankfurt/M, Germany}
\author{C.Krellner}
\address{Physikalisches Institut, Goethe-Universit\"at Frankfurt/M, 60438 Frankfurt/M, Germany}
\date{\today}

\begin{abstract}

Single crystals of HoIr$_2$Si$_2$ with the body-centered ThCr$_2$Si$_2$-type structure ($I4/mmm$) were grown by Bridgman method from indium flux. Single crystal structure determination yielded a Si-z position of 0.378(1) in the structure.
We excluded the presence of the high temperature phase with the primitive CaBe$_2$Ge$_2$-type structure ($P 4/n m m$) by powder X-ray diffraction.
Magnetic measurements on the single crystals yield a \neel temperature of $T_{\rm N}=22\,\rm K$. 
In the inverse magnetic susceptibility a strong anisotropy with Weiss temperatures $\Theta_{W}^{001}=26\,\rm K$ and $\Theta_{W}^{100}=-26\,\rm K$ occurs above $T_{\rm N}$. The effective magnetic moment 
$\mu_{\rm eff}^{001}=10.64\mu_{B}$ and $\mu_{\rm eff}^{100}=10.53\mu_{B}$ is close to the expected value for a free Ho$^{3+}$ ion, $\mu_{\rm eff}^{calc}=10.6\mu_{B}$. 
The field dependent magnetization shows a step-like behaviour due to crystalline electric field effects.


\end{abstract}

\maketitle


\section{Introduction}
Polymorphism in LnIr$_2$Si$_2$ (Ln = lanthanide) is well documented for a large number of compounds, among them for instance CeIr$_2$Si$_2$, YbIr$_2$Si$_2$, LaIr$_2$Si$_2$, GdIr$_2$Si$_2$ and HoIr$_2$Si$_2$ \cite{Niepmann2001, Villars1991, Parthe1984}.
In the case of HoIr$_2$Si$_2$, besides the existence of polymorphism in the body-centered ThCr$_2$Si$_2$-type structure ($I4/mmm$) which is the low temperature phase (LTP) and the primitive CaBe$_2$Ge$_2$-type structure ($P 4/n m m$) which is the high temperature phase (HTP) \cite{Zhong1985}, no characterization of its physical properties was reported in the past. We therefore characterized the magnetic ground state of the LTP of this material. 
In contrast to LnIr$_2$Si$_2$ compounds with the Ln ions Pr \cite{Mihalik2010}, Nd \cite{Welter2003}, Tb \cite{Slaski1983}, Dy \cite{Melamud1984} or Er \cite{Sanchez1993} (Tab.~\ref{Magneticorder}), the study of Ho compounds by neutron scattering experiments is hindered due to its high absorption-cross section for neutrons. We therefore obtained information on the magnetic properties through bulk magnetization measurements.
No successful single crystal growth of HoIr$_2$Si$_2$
had been reported in the literature so far. 
We have grown HoIr$_2$Si$_2$ single crystals from indium flux and present their characterization in this manuscript. 
The relative $z$ position of the Si atoms was found to have a strong impact on the band structure and the Fermi surface topology in the ThCr$_2$Si$_2$-type structure \cite{Reiss2013}. We determined the single crystal structure and the Si-z position in HoIr$_2$Si$_2$ to give reliable input for band structure calculations.

\begin{table}
\centering
\begin{tabular}{|c|c|c|c|c|c|c|}
\hline
    &  $T_{\rm N}$   &$\vec{k}$                  &$T_{\rm N2}$    &$\vec{k}$&moment &ref.\\[+0.5em]
            &  $ [\rm K]$    &  $T_{\rm N2}<T<T_{\rm N}$ &$ [\rm K]$  &      $T<T_{\rm N2}$       &orientation&\\[+0.1em]
\hline\hline
PrIr$_2$Si$_2$& 45.5          &$(0\, 0\, 5/6)$        &$23.7$         &$(0 0 1) $ &  $-$  &\cite{Mihalik2010}\\[+0.1em]
\hline
NdIr$_2$Si$_2$& 33          &   $-$    &$18$         &$(0\, 0\, 1) $ &  $ \parallel c$   &\cite{Welter2003}\\[+0.1em]
\hline
TbIr$_2$Si$_2$& $72\pm 3$          &  $(0\, 0\, 1)$     &$-$         &$-$ & $\parallel c$   &\cite{Slaski1983}\\[+0.1em]
\hline
DyIr$_2$Si$_2$& $40\pm 3$          &  $(0\, 0\, 1)$     &$-$         &$-$ & $\parallel c$   &\cite{Melamud1984}\\[+0.1em]
\hline
ErIr$_2$Si$_2$& $10\pm 3$          &  $(0\, 0\, 1)$     &$-$         &$-$ &  $\perp c$   &\cite{Sanchez1993}\\[+0.1em]
\hline
\end{tabular}
\caption[LnIr$_2$Si$_2$: Magnetic order]
{LnIr$_2$Si$_2$ ($I4/mmm$): Magnetic order below $T_{\rm N}$ with magnetic propagation vector $\vec{k}$ determined by neutron diffraction. In some compounds, a second magnetically ordered phase was found below $T_{\rm{N2}}$. \label{Magneticorder}}
\end{table}

\section{Experiment}
\subsection{Crystal growth}
Single crystals of HoIr$_2$Si$_2$ were grown in indium flux following the route described earlier for the related compounds YbRh$_2$Si$_2$ \cite{Krellner2012a} and GdRh$_2$Si$_2$ \cite{Kliemt2015}. 
The high purity 
starting materials Ho (99.9\%, Strem Chemicals), Ir (99.96\%, Heraeus), 
Si (99.9999\%, Wacker) 
and In (99.9995\%, Schuckard) were weighed 
in a graphite crucible and sealed in a niobium crucible 
under argon atmosphere (99.999\%). 
The initial weight of Ho, Ir and Si was chosen to be stoichiometric and an elements-flux ratio of (Ho,Ir,Si) : In = (4 : 96) at\% was used. 
All experiments have been performed in a movable high-temperature Bridgman furnace (GERO HTRV 70-250/18) where the temperature was measured at the bottom of the niobium crucible using a thermocouple type B. 
The furnace was heated up to the maximum temperature of $T_{\rm max}=1550^{\circ}$C with a rate of $300\,\rm K/h$.
After a homogenization period of $1\,\rm h$ the furnace was moved upwards 
applying a fast-move period ($v = 100\,\rm mm/h$) in the first 
and a slow-move period ($v = 1\,\rm mm/h$) in the second step.
The first step was done to lower the temperature quickly to $T_{\rm 1}=1535^{\circ}\rm C$
after the homogenization of the melt. This avoids the long exposure 
at a high temperature which might lead to an enhanced pollution of the melt 
by the crucible material graphite.
During the slow-move period, the crystal growth took place. 
This step was terminated at $T_{\rm 2}=1090^{\circ}\rm C$. Afterwards, the furnace was slowly cooled down with $50\,\rm K/h$ to room temperature. 
The slow cooling rate was chosen to include an annealing period and to obtain crystals in the low-temperature ThCr$_2$Si$_2$-type crystal structure $I4/mmm$.
Subsequently, the indium flux was removed from the single crystals by etching using 32$\%$ hydrochloric acid.
This growth procedure yielded single crystal platelets of 2$\times$3\,mm$^2$ and a thickness of up to $650\,\mu\rm m$ with masses up to $30\,\rm mg$. Fig.~\ref{HoIr2Si2rho} (d), shows a typical single crystal.

\subsection{Characterization}
Powder X-ray diffraction (PXRD) data were collected
with a Bruker D8 diffractometer with CuK$_{\alpha}$ radiation ($\lambda = 1.5406$\,\AA) at room temperature.
PXRD on crushed single crystals of HoIr$_2$Si$_2$ 
confirmed the $I4/mmm$ tetragonal structure and yielded lattice constants which are summarized in Tab.~\ref{HoIr2Si2_Littabelle}. 
For the single crystal structure determination the 
data were collected at $173\,\rm K$ on a STOE IPDS II two-circle diffractometer 
with a Genix Microfocus tube with mirror optics using MoK$_{\alpha}$ radiation 
($\lambda = 0.71073$\,\AA) and were scaled using the frame scaling procedure 
in the X-AREA program system \cite{Stoe2002}. 
The structure was solved by direct methods using the program SHELXS \cite{Sheldrick2008}
and refined against $F^2$ with full-matrix least-squares techniques using the program 
SHELXL \cite{Sheldrick2008}.
The atomic coordinates determined by single crystal analysis are summarized in Tab.~\ref{HoIr2Si2_AtomicCoord}.
Energy dispersive X-ray spectroscopy (EDX) measurements have been performed with an AMETEK EDAX
Quanta400 Detector in a Zeiss scanning electron microscope (SEM) DSM 940A.
The chemical analysis by EDX microprobe analysis confirmed the presence of Ho, Ir and Si in the single crystals. A quantitative analysis was not possible due to the overlap of the Si-K line and the M lines of Ho and Ir.
The orientation of the single crystals was determined 
using a Laue device with 
tungsten anode. The analysis of 15 samples yielded that the dimension perpendicular to the surface is the $[001]$ direction of the tetragonal lattice. The largest naturally grown edges point towards the $[100]$ direction in HoIr$_2$Si$_2$ as shown in Fig.~\ref{HoIr2Si2rho} (d).
Four-point resistivity, heat capacity and magnetization measurements were performed between $1.8\,\rm K$ and $380\,\rm K$
using a Quantum Design Physical Property Measurement System (PPMS).


\begin{table}
\centering
\begin{tabular}{|c|c|c|c|}
\hline
a   &c    &sample           &ref.\\
$[$\AA$]$ &$[$\AA$]$        &            & \\
\hline
4.042 & 9.707&     PC&\cite{Zhong1985}\\
\hline          
4.0371  & 9.8769 &PC    &this work\\
\hline    
4.0476(6)  &9.884(2) &SC    &this work\\
\hline    
\end{tabular}
\caption[HoIr$_2$Si$_2$: Lattice parameters]
{Lattice parameters determined from polycrystalline (PC) and single crystalline (SC) HoIr$_2$Si$_2$ ($I4/mmm$). \label{HoIr2Si2_Littabelle}}
\end{table}

\begin{table}
\centering
\begin{tabular}{|c|c|c|c|c|c|}
\hline
atom  & Wyckoff position & x &y  & z     & U(eq) $[$\AA$^2]$\\
\hline
Ho&  2a     &0  &0  & 0     &0.0283(10)\\
Ir&  4d     &0  &0.5& 0.25  &0.0276(10)\\
Si&  4e     &0  &0  & 0.378(1)&0.029(2)\\
\hline 
\end{tabular}
\caption[HoIr$_2$Si$_2$: Atomic coordinates]
{Atomic coordinates and equivalent isotropic displacement parameters
for HoIr$_2$Si$_2$ ($I4/mmm$) determined by single crystal analysis at $T=173\,\rm K$. U(eq) is defined as one third of the trace
of the orthogonalized $U^{ij}$ tensor. \label{HoIr2Si2_AtomicCoord}}
\end{table}

\section{Results and discussion}

\subsection{Temperature and field dependence of the magnetization}
\begin{figure}
\centering
\includegraphics[width=1.0\textwidth]{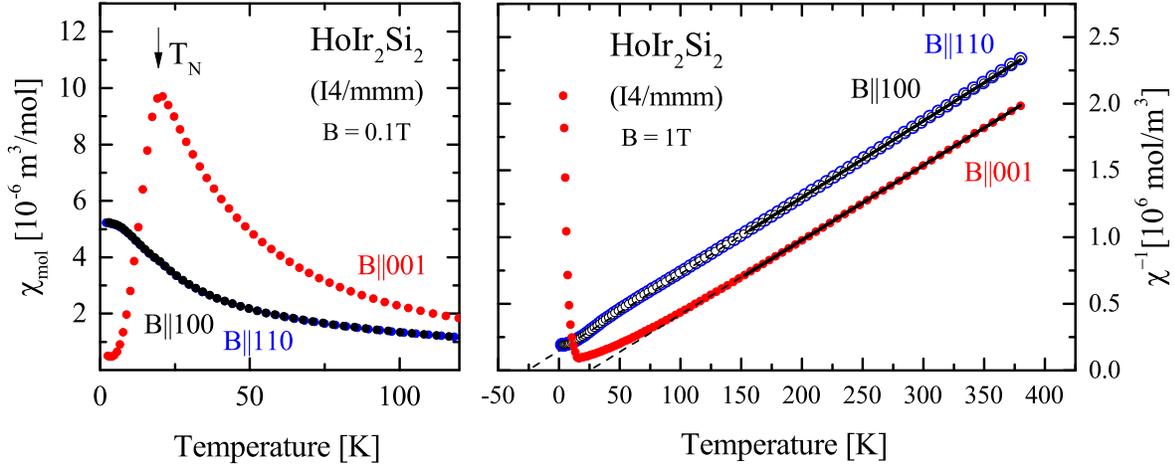}
\caption{{\it Left:} Comparison of the temperature dependence of the susceptibility for an applied field of $B=0.1\,\rm T$ for 3 different crystal orientations. {\it Right:} Inverse susceptibility for $B= 1\,\rm T$. The effective magnetic moments and the Weiss temperatures were determined from the fit to the data between 200 and $380\,\rm K$.}
\label{HoIr2Si2_MvT_Chi_inv}
\end{figure}

Temperature dependent magnetic measurements on the single crystals, Fig.~\ref{HoIr2Si2_MvT_Chi_inv}, left panel,
show the ordering of the Ho$^{3+}$ moments at $T_N=22\,\rm K$. 
Upon increasing the temperature up to the \neel temperature, the moments fluctuate between the up and the down state which leads to an increase of the susceptibility.
We determined the effective magnetic moments and Weiss temperatures from a linear fit between $200$ and $380\,\rm K$ to the inverse magnetic susceptibility as demonstrated in Fig.~\ref{HoIr2Si2_MvT_Chi_inv}, right panel.
The effective magnetic moments 
$\mu_{\rm eff}^{100} = (10.53\pm 0.10)\mu_B$, 
$\mu_{\rm eff}^{110} = (10.52\pm 0.10)\mu_B$ and
$\mu_{\rm eff}^{001} = (10.64\pm 0.10)\mu_B$  
agree well with the predicted 
value of $\mu^{\rm calc}_{\rm eff} = 10.6\,\mu_{\rm B}$ for the free Ho$^{3+}$. 
The Weiss temperatures for field in the $a-a$ plane 
$\Theta_{\rm W}^{100} = (-26\pm 1)\,\rm K$ and 
$\Theta_{\rm W}^{110} = (-26\pm 1)\,\rm K$ 
are isotropic. With $\Theta_{\rm W}^{001} = (26\pm 1)\,\rm K$ a strong anisotropy occurs for field parallel and perpendicular to the $c$-direction. While $\Theta_{W}$ hints to the presence of ferromagnetic fluctuations for field along the $[001]$ direction, antiferromagnetic fluctuations are present for field perpendicular to this direction.

\begin{figure}
\centering
\includegraphics[width=0.49\textwidth]{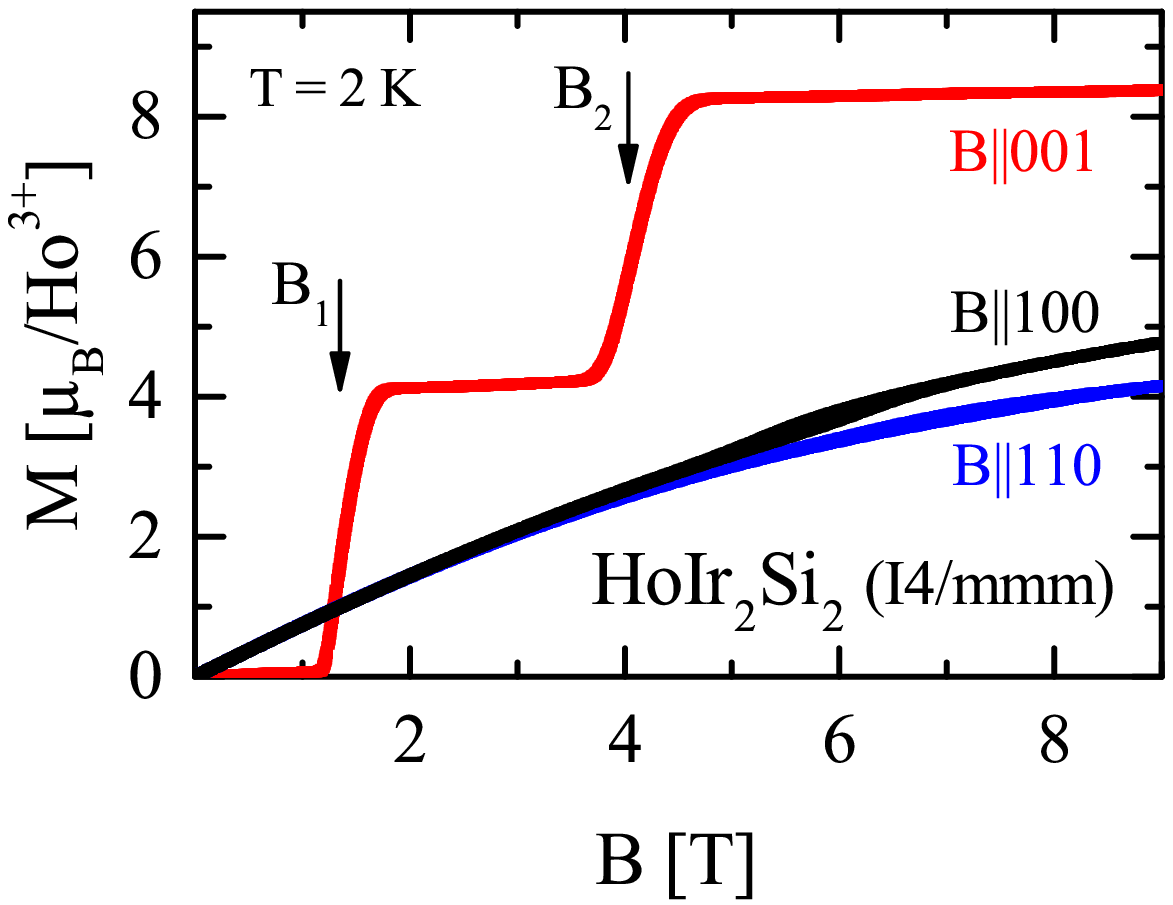}
\includegraphics[width=0.46\textwidth]{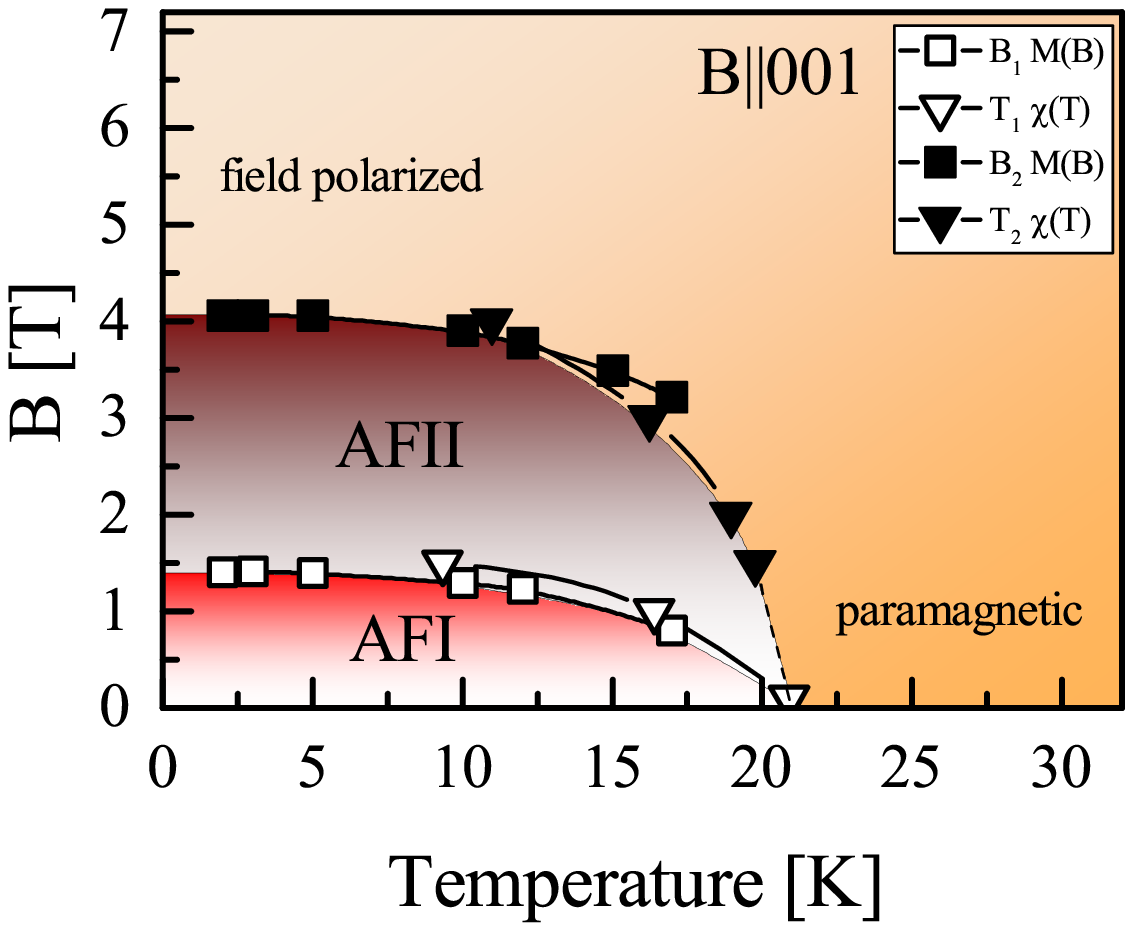}
\caption{{\it Left:} Magnetization $M(B)$ at $T = 2\,\rm K$ for field up to $9\,\rm T$ in 3 different crystal orientations. {\it Right:} $B-T$ magnetic phase diagram for field 
$B\parallel 001$. Closed and open squares were determined from 
the field dependence of the magnetization. 
$B_{c1}$ and $B_{c2}$ are the critical fields 
of the metamagnetic transitions.
Triangles indicate maxima in the temperature
 dependent susceptibility $\chi(T)$.}
\label{HoIr2Si2_MvH}
\end{figure}

The field dependent magnetization within the ordered phase, Fig.~\ref{HoIr2Si2_MvH}, left panel, shows a strong anisotropy for 
$B\parallel 001$ and $B\perp 001$. While for both in-plane directions, the magnetization increases nearly linearly, the susceptibility is nearly zero for $B\parallel 001$ in low fields. 
At $T=2\,\rm K$, a metamagnetic transition occurs at $B\approx 1.4\,\rm T$ and a second one at $B\approx 4.1\,\rm T$ for $B\parallel 001$. This step-like behaviour is caused by the CEF anisotropy.
For $B\parallel 001$, the magnetization reaches $M(9\,\rm T)=8.38\,\mu_{\rm B}/{\rm Ho}^{3+}$ which is slightly smaller than the saturation magnetization $M_{\rm sat}= g_J J=5/4\cdot 8\,\mu_{\rm B}/{\rm Ho}^{3+}= 10\,\mu_{\rm B}/{\rm Ho}^{3+}$.
For $B\parallel 001$ a step-like magnetization occurs while 
for $B\parallel 100$ and $B\parallel 110$ no indications for 
field-induced phases were found.
The $B-T$ magnetic phase diagram for 
$B\parallel 001$ shown in Fig.~\ref{HoIr2Si2_MvH}, right panel, 
was constructed from the field and temperature dependence of the magnetization. 
We found two field induced phases AFI at low fields and AFII at high fields.
The magnetization and susceptibility data, Figs.~\ref{HoIr2Si2_MvT_Chi_inv} and \ref{HoIr2Si2_MvH}, strongly suggest that the magnetic moments order along the $c$ direction. 

\subsection{Heat capacity and entropy}

The heat capacity of HoIr$_2$Si$_2$, shown in Fig.~\ref{HoIr2Si2HC} (a), exhibits a pronounced anomaly at  $T_{\rm N}=22\,\rm K$, 
establishing a second order phase transition into the AFM ordered phase. 
The magnetic part of the heat capacity was determined by subtraction of the data of the non-magnetic reference compound from the HoIr$_2$Si$_2$ data according to $C^{4f}=C$(HoIr$_2$Si$_2$)$- C$(LuIr$_2$Si$_2$).
The magnetic part of the entropy, $S^{4f}$, shown in Fig.~\ref{HoIr2Si2HC} (b), was determined by integration of $C^{4f}/T$.  It reaches $R\,\rm ln 4$ at $T_{\rm N}$ which is much smaller than the 
expected $R\,{\rm ln}(2J+1)=R\,{\rm ln}17$ for the Ho$^{3+}$ ion 
indicating that the overall splitting of the crystal electric field (CEF)
levels is much larger than $T_{\rm N}$. The increase from $R\,{\rm ln}4$ to $R\,{\rm ln}12$ above the \neel temperature to about $100\,\rm K$ can be attributed to the population of excited CEF levels. The entropy saturates at about $R\,{\rm ln}12$ which is far below $R\,{\rm ln}17$. This discrepancy is probably caused by the large uncertainty in the measured heat capacity data for $T>50\,\rm K$.

\subsection{Electrical transport}

\begin{figure}
\centering
\includegraphics[width=1.0\textwidth]{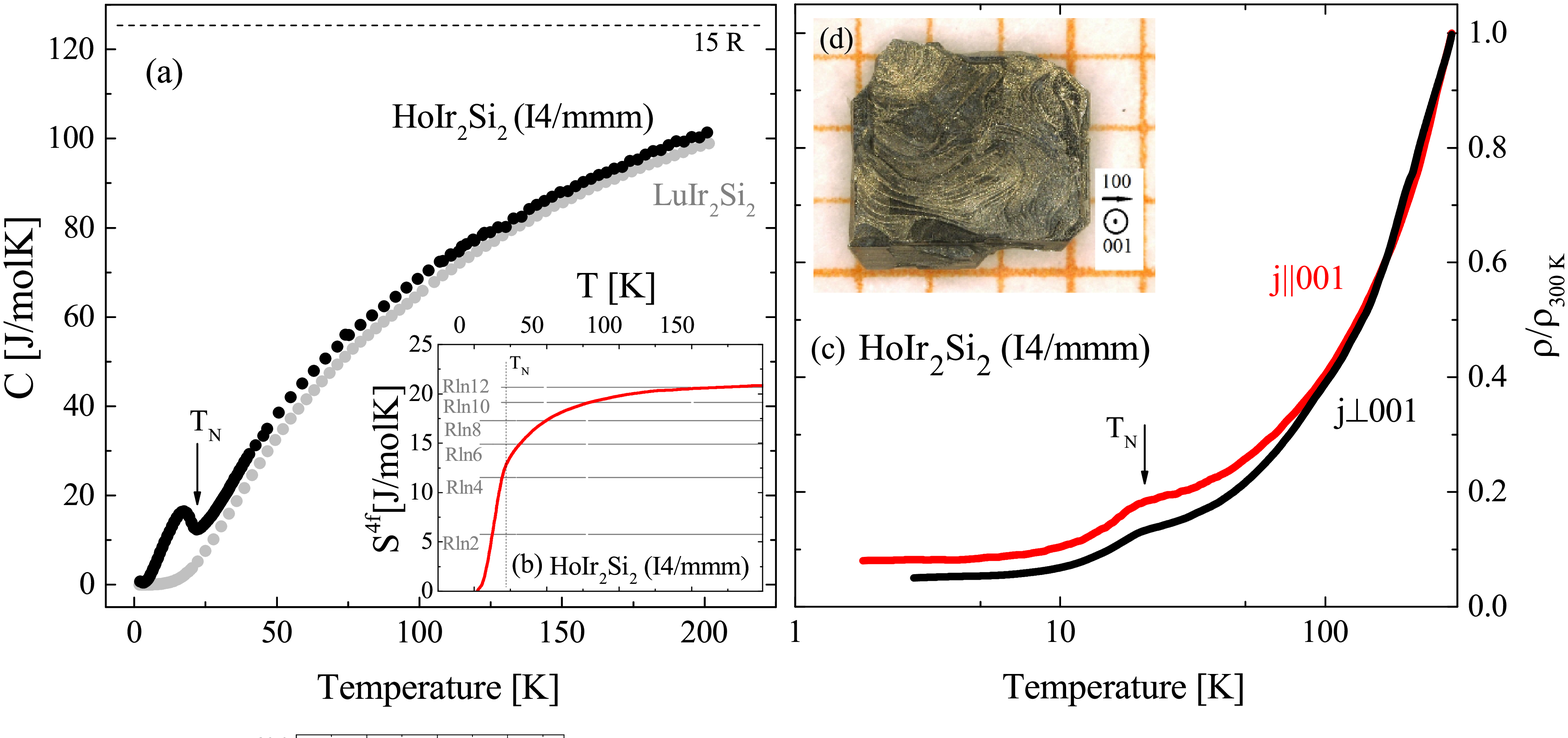}
\caption[HoIr$_2$Si$_2$: Heat capacity $C(T)$ and resistivity $\rho(T)$]{HoIr$_2$Si$_2$ (a) Temperature dependence of the heat capacity. LuIr$_2$Si$_2$ data from \cite{Krellner2009}, (b) Magnetic contribution to the entropy, (c) Temperature dependence of the electrical resistivity. A kink in the curves for both current directions occurs at the \neel temperature, (d) HoIr$_2$Si$_2$ single crystal ($I4/mmm$).}
\label{HoIr2Si2HC}\label{HoIr2Si2rho}
\end{figure}

Electrical transport data, Fig.~\ref{HoIr2Si2rho} (c), show the magnetic transition at $T\approx 21\,\rm K$ and a slight anisotropy for current flow parallel and perpendicular to the $[001]$-direction. 
With $\rm RRR=\rho_{300K}/\rho_{0}\sim 25$ for $j\perp 001$ and $\rm RRR=\rho_{300K}/\rho_{0}\sim 12$ for $j\parallel 001$ the residual resistivity ratio is also anisotropic. 
Upon cooling below the \neel temperature,
a drop of the resistivity is observed for both current directions. 

\section{Summary}

In this work, single crystals of HoIr$_2$Si$_2$ have been grown by a modified Bridgman method from indium flux. After an optimization of the temperature-time profile of the growth experiment, we obtained millimetre-sized single crystals. They have a platelet habitus, with the $c$-axis perpendicular to this platelet and formed in the $I4/mmm$ tetragonal structure, which is the low-temperature phase. Powder X-ray diffraction (PXRD) on crushed single crystals
confirmed this $I4/mmm$ tetragonal structure with lattice parameters which
are in agreement with the data published for polycrystalline samples. The PXRD pattern did not show any additional reflections belonging to the $P 4/n m m$ structure. 
We additionally performed a single crystal structure analysis and determined the Si position to $z=0.378(1)$ in HoIr$_2$Si$_2$ to give reliable input for the band structure calculations. 
We present a detailed study of the bulk properties, namely heat capacity, magnetization and electrical transport of HoIr$_2$Si$_2$ single crystals. 
Our heat capacity measurements confirm the occurence of a second order phase transition at the \neel temperature $T_{\rm N}=22\,\rm K$. 
The temperature and field dependence of the magnetization hints to the ordering of the magnetic moments along the $c$ direction.
In the inverse magnetic susceptibility a strong anisotropy with Weiss temperatures $\Theta_{W}^{001}=26\,\rm K$ and $\Theta_{W}^{100}=-26\,\rm K$ occurs for field parallel and perpendicular to the $[001]$ direction. The effective magnetic moment 
$\mu_{\rm eff}^{001}=10.64\mu_{\rm B}$ and $\mu_{\rm eff}^{100}=10.53\mu_{\rm B}$ is close to the expected value for a free Ho$^{3+}$ ion ($\mu_{\rm eff}^{\rm calc}=10.6\mu_{\rm B}$).
The residual resistivity ratio $\rm RRR=\rho_{300K}/\rho_{0}\sim 25$ for $j\perp 001$ and $\rm RRR=\rho_{300K}/\rho_{0}\sim 12$ for $j\parallel 001$ is anisotropic. 
Furthermore, it is important to determine the magnetic ground state of this material, as it has been shown, that the LnT$_2$Si$_2$ compounds bear intriguing surface states, which are strongly influenced by the magnetic properties of the 4$f$ electrons \cite{Guettler2016, Generalov2017}.
An angle resolved photoemission (ARPES) study on HoRh$_2$Si$_2$ revealed that the temperature can be used as a tuning
parameter for the surface magnetism since the temperature dependent changes in the moment orientation in the bulk reflect in the surface properties \cite{Generalov2017}. 
The comparison of the magnetic interactions occuring in the Rh and the Ir compounds will be a task for future work.
Further studies of HoIr$_2$Si$_2$ to investigate its magnetic structure using resonant inelastic X-ray scattering (RIXS) \cite{Kliemt2017} or electron spin resonance (ESR) \cite{Sichelschmidt2016, Sichelschmidt2017b} are in progress.


\section{Acknowledgments}
We acknowledge discussions with C.Geibel, D.V.Vyalikh and J.Sichelschmidt, as well as funding by the DFG through grant KR3831/5-1.




\section*{References}
\providecommand{\newblock}{}

\end{document}